\documentclass[pra,twocolumn,showpacs,superscriptaddress,floatfix, nofootinbib]{revtex4-2}

\usepackage[english]{babel}
\usepackage[normalem]{ulem}

\usepackage{graphicx}
\usepackage{amsmath,amssymb,mathrsfs,esint}
\usepackage{multirow}
\usepackage{algorithm}
\usepackage[noend]{algpseudocode}
\usepackage{comment}
\usepackage{physics}
\usepackage{bbold}
\usepackage{tabularx}
\usepackage{subcaption}

\DeclareMathOperator{\Span}{Span}
\DeclareMathOperator{\diag}{diag}

\usepackage{bm}
\usepackage{relsize}

\usepackage[normalem]{ulem}
\usepackage[bookmarks=true,
   colorlinks=true,
   linkcolor=blue,
   urlcolor=blue,
   citecolor=blue,
   bookmarks=true,
   hyperindex=true
]{hyperref}

\usepackage{tikz}
\usetikzlibrary{quantikz2}

\DeclareUnicodeCharacter{00A0}{~}

\begin{document}

\title{On the local equivalence of trapped-ion two-qudit gates}

\author{Nikita V. Semenin}\email{semeninnv@gmail.com}
\affiliation{P.N. Lebedev Physical Institute of the Russian Academy of Sciences, Moscow 119991, Russia}

\author{Pavel A. Kamenskikh}
\affiliation{P.N. Lebedev Physical Institute of the Russian Academy of Sciences, Moscow 119991, Russia}

\author{Ilia V. Zalivako}
\affiliation{P.N. Lebedev Physical Institute of the Russian Academy of Sciences, Moscow 119991, Russia}

\author{Anastasiia S. Nikolaeva}
\affiliation{P.N. Lebedev Physical Institute of the Russian Academy of Sciences, Moscow 119991, Russia}
\affiliation{Russian Quantum Center, Skolkovo, Moscow 121205, Russia}

\author{Evgeniy O. Kiktenko}
\affiliation{
Steklov Mathematical Institute of Russian Academy of Sciences, Gubkina St.
8, Moscow 119991}
\affiliation{Russian Quantum Center, Skolkovo, Moscow 121205, Russia}

\begin{abstract}
We derive a necessary condition of the local equivalence between two-qudit gates in terms of singular values of transformed gate matrices. This condition is valid for arbitrary qudit dimensions~$d$ and is thus a relatively simple general way of checking whether two gates can be reduced to one another with single-qudit (local) gates. We use this condition to investigate the local equivalence of two widely used trapped-ion two-qubit gates in qudit space: the M\o lmer-S\o rensen (MS) gate and a special case of the Light-Shift (LS) gate, both of which we studied in one of our previous works.
\end{abstract}

\maketitle

\section{Introduction}
An important quality of two entangling quantum gates between two qudits is whether they are locally equivalent, that is, whether they can be transformed into one another by using only local (single-qudit) gates. For qubits, there has been a substantial amount of research on this topic (see~\cite{makhlin2002nonlocal, zhang2003geometric} and references therein), which cannot quite be said for qudits. With qudit systems becoming more prominent in the field of quantum computing, there is a high demand for thorough theoretical analysis of qudit gates, especially compared to their qubit counterparts.

In our previous work~\cite{kamenskikh2026analysis}, we studied the extension of conventional trapped-ion two-qubit gates to qudits. Namely, the M\o lmer-S\o rensen (MS) gate and the Light-Shift (LS) gate, the ideal evolution operators for which can be expressed as (up to an equivalent redefinition of the variables):
\begin{gather}
\label{eq:U_MS}
U_{\mathrm{MS}} =  \exp\left(i \chi \sigma_{x}^{01} \sigma_{x}^{01} + i  \sum_{j = 1}^2\sum\limits_{s=0}^{d-1}\phi_{j,s}^{\mathrm{MS}}\ketbra{s_j}{s_j}\right),\\
\label{eq:U_LS}
\begin{multlined}
U_{\mathrm{LS}} = \exp\left( i \sum_{s,s' = 0}^{d-1} \chi_{ss'} \ketbra{ss'}{ss'}\right.\\
\left.+i \sum_{j = 1}^2\sum\limits_{s=0}^{d-1}\phi_{j,s}^{\mathrm{LS}}\ketbra{s_j}{s_j}
\right),
\end{multlined}
\end{gather}
where
\begin{equation}
\sigma_x^{jk}=\ketbra{j}{k} + \ketbra{k}{j}
\end{equation}
is the Pauli matrix extension for qudits, and~$\phi_{j,s}^{\mathrm{MS}},\phi_{j,s}^{\mathrm{LS}}$ are the local phases which are induced by the interaction with the motional modes of the ionic chain and by the laser Stark shifts of the basis states. In this note, we restrict our analysis to the \emph{zero-order} LS gate, which has a single nonzero nonlocal phase~$\chi_{00}\equiv\chi$. The reason for this restriction is two-fold. First, the zero-order LS gate is the simplest variation of the general LS gate that depends on only one nonlocal parameter, the same number of nonlocal parameters as the MS gate, allowing them to be compared in terms of their local equivalence. Second, the zero-order case takes place when only one of the qudit states is strongly coupled to the laser field driving the gate. Such configurations are already used in experimental LS gate implementations for ionic qubits~\cite{baldwin_high-fidelity_2021} and qudits~\cite{hrmo_native_2023}, further motivating this consideration. Henceforth, we will assume the zero-order LS gate.

In the present note, we prove the necessary condition that any two two-qudit gates must satisfy to be locally equivalent. The condition is based on the comparison between the singular value spectra of the qudit gate matrices transformed in a specific way. We then use this condition to see whether the MS and zero-order LS gates are locally equivalent. The final section concludes our findings.
\section{The necessary condition for the equivalence of two-qudit gates}
Let $\mathcal{H}_A$ and $\mathcal{H}_B$ be Hilbert spaces associated with two
finite-dimensional quantum systems, with $\dim \mathcal{H}_A=d_A$,
$\dim \mathcal{H}_B=d_B$.
We denote by $\mathcal{U}(\mathcal{H}_A \otimes \mathcal{H}_B)$ the group of
unitary operators acting on the composite Hilbert space
$\mathcal{H}_A \otimes \mathcal{H}_B$. Let
$U,V \in \mathcal{U}(\mathcal{H}_A \otimes \mathcal{H}_B)$
be two bipartite unitary operators.
We say that $U$ and $V$ are locally equivalent if there exist local unitary
operators $
A_1,A_2 \in \mathcal{U}(\mathcal{H}_A)$, $B_1,B_2 \in \mathcal{U}(\mathcal{H}_B)$,
such that
\begin{equation}
V=(A_1\otimes B_1)\,U\,(A_2\otimes B_2).
\end{equation}

\begin{figure*}[t]
    \centering
    \includegraphics[width=0.95\linewidth]{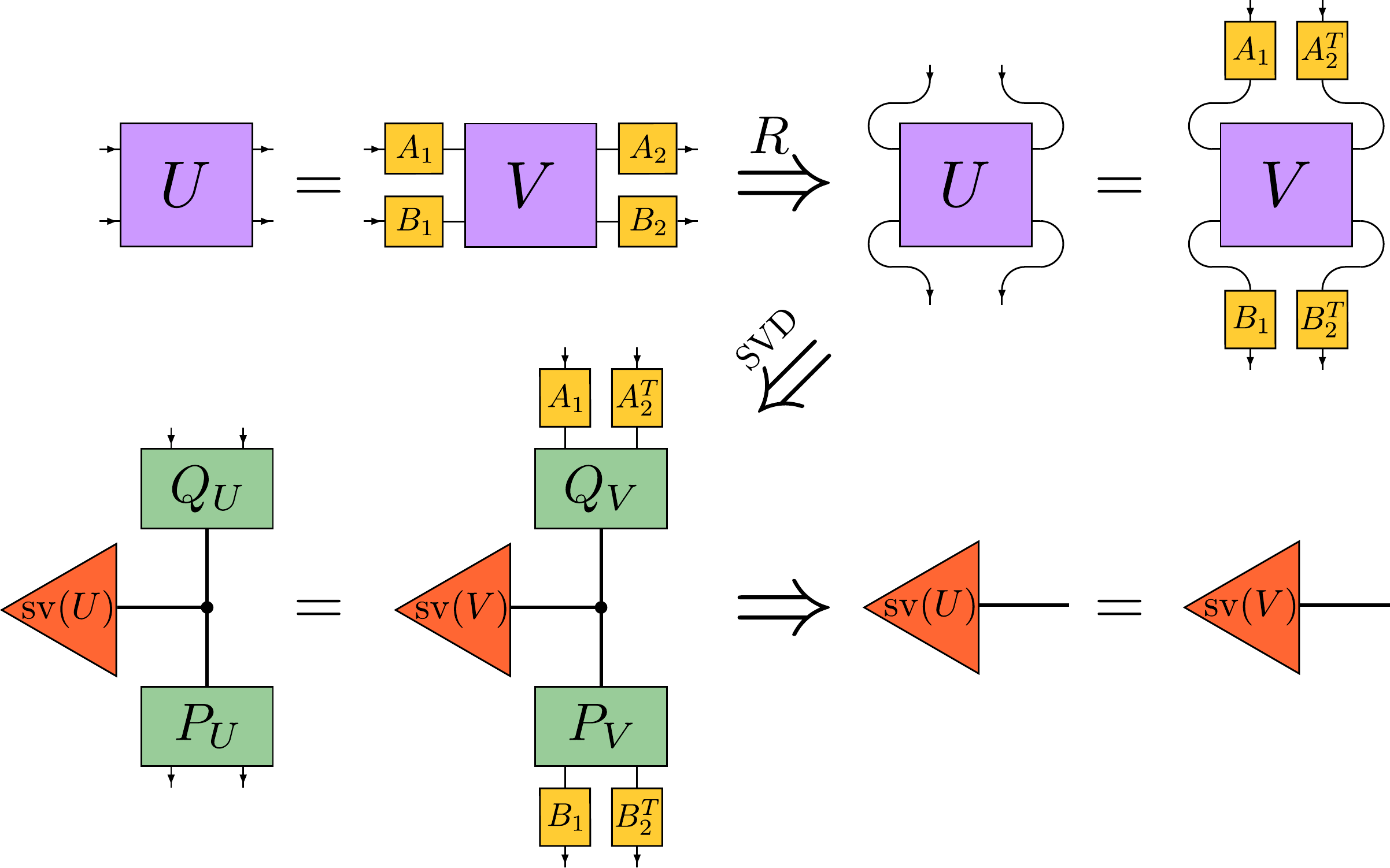}
    \caption{Schematic proof of the \textbf{Theorem} on the local equivalence of two-qudit gates using tensor network notation. $Q_X$ and~$P_X$ denote the left and right unitary matrices obtained from the singular value decomposition of the matrix~$X$.  }
    \label{fig:equivalence}
\end{figure*}

Let the matrix elements of $U$ be written as
\begin{equation}
U_{ij,kl}=\langle i,j|U|k,l\rangle,
\end{equation}
where the indices $i,k$ correspond to subsystem $A$, while $j,l$
correspond to subsystem $B$. We define the reshuffled matrix $R(U)$ as the
$d_A^2\times d_B^2$ matrix with elements
\begin{equation}
\label{eq:reshuffling_transform}
R(U)_{ik,jl}=U_{ij,kl},
\end{equation}
where the pairs of indices $(i,k)$ and $(j,l)$ are treated as composite
(super-)indices labeling the rows and columns of $R(U)$, respectively. See Figure~\ref{fig:equivalence} for visual explanation of the reshuffling operator~$R$.
The singular values of $R(U)$ are referred to as the operator Schmidt
coefficients of $U$.

\textbf{Theorem.}
Let $\operatorname{sv}(X)$ denote the set of singular values of a matrix $X$,
counted with multiplicities. If two bipartite unitary operators $U$ and $V$
are locally equivalent, then their operator Schmidt coefficients coincide:
\begin{equation}
\operatorname{sv}(R(U))=\operatorname{sv}(R(V)).
\end{equation}
Consequently, if $\operatorname{sv}(R(U))\neq \operatorname{sv}(R(V))$,
then $U$ and $V$ are not locally equivalent.

\textbf{Proof.}
Assume that
\begin{equation}
V=(A_1\otimes B_1)\,U\,(A_2\otimes B_2).
\end{equation}
Then, under the reshuffling map defined above (see also Fig.~\ref{fig:equivalence}), the corresponding reshuffled
matrices are related as
\begin{equation}
R(V)=(A_1\otimes A_2^{T})\,R(U)\,(B_1^{T}\otimes B_2),
\end{equation}
up to the chosen convention for ordering the indices. Since the matrices
multiplying $R(U)$ from the left and from the right are unitary, the singular
values of $R(U)$ are invariant under this transformation. Therefore, locally
equivalent operators have the same set of operator Schmidt coefficients.
\begin{equation}
\operatorname{sv}(R(U))=\operatorname{sv}(R(V)).
\end{equation}
This proves the statement. \(\square\)

Thus, a mismatch between the operator Schmidt spectra is a sufficient witness
of inequivalence with respect to local unitary transformations. However, the
coincidence of these spectra alone does not, in general, imply local
equivalence.

As an example, consider the two-qubit gates
\begin{equation}
\mathrm{SWAP} =
\begin{pmatrix}
1&0&0&0\\
0&0&1&0\\
0&1&0&0\\
0&0&0&1
\end{pmatrix},
\qquad
\mathrm{iSWAP} =
\begin{pmatrix}
1&0&0&0\\
0&0&i&0\\
0&i&0&0\\
0&0&0&1
\end{pmatrix}.
\end{equation}
For the reshuffling convention $R(U)_{ik,jl}=U_{ij,kl}$, with the composite
indices ordered as $(00),(01),(10),(11)$, one obtains
\begin{equation}
R(\mathrm{SWAP}) = \mathrm{SWAP},
\qquad
R(\mathrm{iSWAP}) = \mathrm{iSWAP}.
\end{equation}
Thus, both reshuffled matrices are unitary. Hence all their singular values are
equal to one:
\begin{equation}
\operatorname{sv}(R(\mathrm{SWAP}))
=
\operatorname{sv}(R(\mathrm{iSWAP}))
=
(1,1,1,1).
\end{equation}
Therefore, the operator Schmidt spectra coincide.

However, the two gates are not locally equivalent. Indeed, SWAP maps any
product state to another product state and therefore has zero entangling
power. The same is true for any gate locally equivalent to SWAP, since local
unitaries cannot create entanglement.

By contrast, iSWAP can generate entanglement. For example,
\begin{equation}
\mathrm{iSWAP}\ket{++}
=
\frac{1}{2}
\left(
\ket{00}
+i\ket{01}
+i\ket{10}
+\ket{11}
\right).
\end{equation}
The resulting state is maximally entangled. Hence iSWAP has nonzero
entangling power, whereas SWAP has zero entangling power. Consequently, SWAP
and iSWAP are not locally equivalent, even though their operator Schmidt
spectra coincide.

\section{Study of the equivalence of the MS and the zero-order LS gates for trapped-ion qudits}
Assuming~$d_A = d_B = d$ and ignoring the local phases~$\phi_{j,s}^{\mathrm{MS}},\phi_{j,s}^{\mathrm{LS}}$ in~\eqref{eq:U_MS} and~\eqref{eq:U_LS}, the unitary operators of the zero-order LS and the MS gates are defined as follows:
\begin{equation}
U_{\mathrm{LS}}(\chi) = e^{i\chi\ketbra{00}{00}},\quad U_{\mathrm{MS}}(\chi) = e^{i\chi\sigma_x^{01}\sigma_x^{01}}.
\end{equation}
The LS gate matrix is therefore
\begin{equation}
U_{\mathrm{LS}} = \diag(e^{i\chi},1,\dots,1).
\end{equation}
After the transformation described by~\eqref{eq:reshuffling_transform}, the matrix will take the following form:
\begin{widetext}
\begin{equation}
\label{eq:LS_big}
\widetilde{U}_{\mathrm{LS}} = R(U_{\mathrm{LS}}) = 
\begin{pmatrix}
e^{i\chi} & 0 & \cdots & 0 & 1 & 0 & \cdots & 0 & 1 & 0 & \cdots & \cdots & 0 & 1\\
0 & \cdots & \cdots & \cdots & 0 & \cdots & \cdots & \cdots & 0 & \cdots & \cdots & \cdots & \cdots & 0\\
\vdots &  &  &  & \vdots &  &  &  & \vdots &  &  &  &  & \vdots \\
0 & \cdots & \cdots & \cdots & 0 & \cdots & \cdots & \cdots & 0 & \cdots & \cdots & \cdots & \cdots & 0 \\
1 & 0 & \cdots & 0 & 1 & 0 & \cdots & 0 & 1 & 0 & \cdots & \cdots & 0 & 1\\
0 & \cdots & \cdots & \cdots & 0 & \cdots & \cdots & \cdots & 0 & \cdots & \cdots & \cdots & \cdots & 0 \\
\vdots &  &  &  & \vdots &  &  &  & \vdots &  &  &  &  & \vdots \\
\vdots &  &  &  & \vdots &  &  &  & \vdots &  &  &  &  & \vdots \\
1 & 0 & \cdots & 0 & 1 & 0 & \cdots & 0 & 1 & 0 & \cdots & \cdots & 0 & 1
\end{pmatrix}.
\end{equation}
\end{widetext}
The number of zeros between the nonzero elements in any row or column is~$d$. The squares of singular values are the eigenvalues of the product~$\widetilde{U}_{\mathrm{LS}}^{\dagger}\widetilde{U}_{\mathrm{LS}}$. Using~\eqref{eq:LS_big}, the product can be written as
\begin{equation}
\widetilde{U}_{\mathrm{LS}}^{\dagger}\widetilde{U}_{\mathrm{LS}} =
\begin{pmatrix}
d & \cdots & B & \cdots & B & \cdots & \cdots & B\\
\vdots &  & \vdots &  & \vdots &  &  & \vdots\\
B^* & \cdots & d & \cdots & d & \cdots & \cdots & d\\
\vdots &  & \vdots &  & \vdots &  &  & \vdots\\
B^* & \cdots & d & \cdots & d & \cdots & \cdots & d\\
\vdots &  & \vdots &  & \vdots &  &  & \vdots\\
\vdots &  & \vdots &  & \vdots &  &  & \vdots\\
B^* & \cdots & d & \cdots & d & \cdots & \cdots & d
\end{pmatrix},
\end{equation}
where
\begin{equation}
B = e^{-i\chi} + (d-1).
\end{equation}
By rearranging the rows and columns, we can compress all nonzero elements of this matrix into a~$d\times d$ block 
\begin{equation}
M_{\mathrm{LS}} = 
\begin{pmatrix}
d & B & B & \cdots & B\\
B^* & d & d & \cdots & d\\
B^* & d & d & \cdots & d\\
\vdots & \vdots & \vdots & & \vdots\\
B^* & d & d & \cdots & d
\end{pmatrix},
\end{equation}
which contains all nonzero eigenvalues. For~$M_{\mathrm{LS}}$, we note that the span of vectors
\begin{equation}
v_1=(1,0,\dots,0)^{\mathrm{T}},\quad v_2=(0,1,\dots,1)^{\mathrm{T}}
\end{equation}
is an invariant subspace. Namely,
\begin{equation}
\label{eq:invariant_subspace_LS}
\begin{gathered}
M_{\mathrm{LS}}v_1 = dv_1+B^*v_2,\\
M_{\mathrm{LS}}v_2 = (d-1)Bv_1+ (d-1)dv_2.
\end{gathered}
\end{equation}
Any vector~$w$ orthogonal to this subspace will have the following properties, defined by~$v_1,v_2$:
\begin{equation}
w_1 = 0,\quad\sum\limits_{i\geq 1}w_i=0.
\end{equation}
It is straightforward to see that~$M_{\mathrm{LS}}w=0$, thus allowing us to further restrict the action of~$M_{\mathrm{LS}}$ to~$\Span{\lbrace v_1,v_2\rbrace}$. The restriction matrix of~$M_{\mathrm{LS}}$ in~$v_1,v_2$ basis is obtained using~\eqref{eq:invariant_subspace_LS}:
\begin{equation}
\left.M_{\mathrm{LS}}\right|_{(v_1,v_2)}=
\begin{pmatrix}
d & B(d-1) \\
B^* & d(d-1),
\end{pmatrix}
\end{equation}
the eigenvalues of which and, consequently, the squared nonzero Schmidt coefficients of the reshuffled gate matrix will be
\begin{equation}
\label{eq:LS_Schmidt}
\lambda^2_{\mathrm{LS}}(\chi) = \dfrac{1}{2}\left(d^2 \pm \sqrt{d^4 - (4(d-1)\sin(\chi/2))^2}\right),
\end{equation}
meaning that, for~$d>2$, there will always be only at most two nonzero Schmidt coefficients.

For the MS gate, we use a basis change which transforms the Pauli matrices~$\sigma_x\rightarrow\sigma_z$ (for example, by applying a Hadamard gate between~$\ket{0}$ and~$\ket{1}$ to each qudit before and after the MS), which allows to represent the MS gate as a diagonal matrix
\begin{equation}
U_{\mathrm{MS}} = \diag(e^{i\chi},e^{-i\chi},\underbrace{1,\dots,1}_{d-2},e^{-i\chi},e^{i\chi},1,\dots,1).
\end{equation}
Using the same procedure as for the zero-order LS gate matrix, we arrive at a~$d\times d$ block
\begin{equation}
M_{\mathrm{MS}} = 
\begin{pmatrix}
d & D & C & \cdots & C\\
D & d & C & \cdots & C\\
C & C & d & \cdots & d\\
\vdots & \vdots & \vdots & & \vdots\\
C & C & d & \cdots & d
\end{pmatrix},
\end{equation}
where
\begin{equation}
\begin{aligned}
C&=d-4\sin^2(\chi/2),\\
D&=d-4\sin^2\chi.
\end{aligned}
\end{equation}
The invariant subspace which contains all nonzero eigenvalues is
\begin{equation}
\begin{gathered}
\Span\lbrace u_1,u_2,u_3\rbrace,\\
u_1=(1,0,\dots,0)^{\mathrm{T}},\\
u_2=(0,1,0,\dots,0)^{\mathrm{T}},\\
u_3=(0,0,1,\dots,1)^{\mathrm{T}},
\end{gathered}
\end{equation}
in which basis the restriction will take the form
\begin{equation}
\left.M_{\mathrm{MS}}\right|_{(u_1,u_2,u_3)}=
\begin{pmatrix}
d & D & (d-2)C\\
D & d & (d-2)C\\
C & C & (d-2)d
\end{pmatrix},
\end{equation}
and the eigenvalues of this matrix will be, again, the Schmidt coefficients squared:
\begin{multline}
\lambda_{\mathrm{MS}}^2(\chi)=\left\lbrace4\sin^2\chi,\dfrac{1}{2}\left((d^2-4\sin^2\chi)\vphantom{\dfrac{1}{2}}\right.\right.\\
\left.\left.\pm\sqrt{(d^2-4\sin^2\chi)^2 - (8(d-2)\sin^2(\chi/2))^2}\right)\right\rbrace.
\end{multline}
For~$d>2$, there will be three nonzero Schmidt coefficients, unless~$\chi=\pi$ (we ignore the trivial case of the identity gate matrix for~$\chi=0\ \text{or}\ 2\pi$). Therefore, for~$\chi\neq\pi$, the qudit MS and zero-order LS gates are not locally equivalent. 

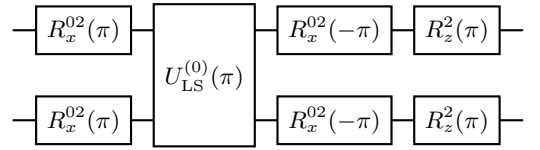
\begin{figure}[h]
\centering
\begin{quantikz}[column sep=0.3cm]
    &\gate{R^{02}_x(\pi)}
    &\gate[wires=2]{U_{\mathrm{LS}}^{(0)}(\pi)}&\gate{R^{02}_x(-\pi)}& \gate{R_z^{2}(\pi)}&
    \\
     &\gate{R^{02}_x(\pi)}& &\gate{R^{02}_x(-\pi)}&\gate{R_z^{2}(\pi)}&
\end{quantikz}
\caption{Decomposition of~$U_{\mathrm{MS}}(\pi)$ into~$U_{\mathrm{LS}}^{(0)}(\pi)$ using local unitaries for~$d=3$}
\label{fig:LS_MS_equiv}
\end{figure}

We now consider the case~$\chi=\pi$ (in the MS gate). In order to satisfy the equivalence condition, we need to find at least one phase~$\varphi$, such that
\begin{equation}
\lambda_{\mathrm{LS}}^2(\varphi) = \lambda_{\mathrm{MS}}^2(\pi).
\end{equation}
For~$d=3$, the only phase that satisfies this condition is~$\varphi=\pi$, in which case
\begin{equation}
\lambda_{\mathrm{LS}}^2(\pi)=\lambda_{\mathrm{MS}}^2(\pi)=\dfrac{1}{2}\left(9\pm\sqrt{17}\right).
\end{equation}
It turns out that, in this specific case, the gates are equivalent. The decomposition of the MS($\pi$) into the zero-order LS($\pi$) for~$d=3$ and local unitaries is demonstrated in Fig.~\ref{fig:LS_MS_equiv}.
The notation for single-qudit rotations is consistent with~\cite{kamenskikh2026analysis}, that is:
\begin{equation}
\begin{gathered}
R_{x}^{jk}(\theta) = \exp\left(- i \frac{\theta}{2}\sigma_{x}^{jk}\right),\\
R_z^{j}(\theta) = \exp( i \theta \ketbra{j}{j}).
\end{gathered}
\end{equation}

For~$d>3$, the Schmidt numbers of the MS gate at~$\chi=\pi$ will be
\begin{equation}
\lambda_{\mathrm{MS}}^2(\pi) = \dfrac{1}{2}\left(d^2 \pm \sqrt{d^4 - (8(d-2))^2}\right).
\end{equation}
At the same time, the greater of the two Schmidt numbers for the LS gate~\eqref{eq:LS_Schmidt} will be no less than its value at~$\pi$:
\begin{equation}
(\lambda_{\mathrm{LS}}^2(\varphi))_{\mathrm{max}} \geq \dfrac{1}{2}\left(d^2 + \sqrt{d^4 - (4(d-1))^2}\right).
\end{equation}
A similar bound can be obtained for the lesser Schmidt number:
\begin{equation}
(\lambda_{\mathrm{LS}}^2(\varphi))_{\mathrm{min}} \leq \dfrac{1}{2}\left(d^2 - \sqrt{d^4 - (4(d-1))^2}\right).
\end{equation}
Since $d>3$, this means that~$8(d-2)>4(d-1)$, therefore, we get the following chain of inequalities:
\begin{multline}
(\lambda_{\mathrm{LS}}^2(\varphi))_{\mathrm{min}} < (\lambda_{\mathrm{MS}}^2(\pi))_{\mathrm{min}} \\
< (\lambda_{\mathrm{MS}}^2(\pi))_{\mathrm{max}} < (\lambda_{\mathrm{LS}}^2(\varphi))_{\mathrm{max}}.
\end{multline}
Thus, the Schmidt numbers of the MS gate at~$\pi$ are always strictly between the Schmidt numbers of the LS gate at any phase, making the equivalence of these gates impossible.
\section{Conclusion}
We have found a necessary condition for the local equivalence of any two-qudit gates, which can be a useful tool for detecting gates that are definitely non-equivalent. In addition, we have proven that the zero-order LS gate and the MS gate in ion qudits are only locally equivalent if~$d=3$ and the nonlocal phases in both gates are~$\chi=\pi$.
\section*{Acknowledgments}
The work of E.O. Kiktenko was performed at the Steklov International Mathematical Center and supported by the Ministry of Science and Higher Education of the Russian Federation (agreement no. 075-15-2025-303).
\bibliography{bibliography.bib}
\end{document}